\documentclass[aps,prx,twocolumn,superscriptaddress,nofootinbib]{revtex4-2}

\usepackage{amsmath,amssymb,amsfonts}
\usepackage{graphicx}
\usepackage{bm}
\usepackage{physics}
\usepackage{braket}
\usepackage[colorlinks=true,allcolors=blue]{hyperref}
\usepackage{mathtools}
\usepackage{microtype}
\usepackage{float} 
\usepackage{placeins}
\newcommand{\Eexact}{E_0}
\newcommand{\Evqe}{E_{\mathrm{VQE}}}
\newcommand{\HCI}{H_{\mathrm{CI}}}
\newcommand{\Htc}{H_{\mathrm{TC}}}

\usepackage{amsthm}

\begin{document}


\title{Separation of Statistical Complexity and Trainability in Variational Quantum Circuits}

\author{Suman Mandal}
\affiliation{Department of Physics, University of Central Florida, Orlando, Florida 32816, USA}

\author{Maximillian Daughtry}
\affiliation{Department of Physics, University of Central Florida, Orlando, Florida 32816, USA}

\author{Eduardo R. Mucciolo}
\affiliation{Department of Physics, University of Central Florida, Orlando, Florida 32816, USA}

\date{June 16, 2026}

\begin{abstract}
Variational quantum algorithms (VQAs) are among the leading approaches for near-term quantum computing, yet their performance can degrade in barren plateau regimes characterized by vanishing gradients. A widely held intuition is that increasing circuit expressivity, often associated with random-state behavior, leads to a loss of trainability. Existing results show that sufficiently random circuits can lead to barren plateaus. Here we show that standard statistical signatures of randomness can emerge well before this regime, without degrading trainability. We demonstrate this behavior in structured variational circuits applied to the one-dimensional cluster-Ising model and a generalized toric code Hamiltonian. To characterize state complexity, we analyze Porter-Thomas statistics, entanglement-spectrum level statistics, and inverse participation ratios. Across both models, increasing circuit depth drives these diagnostics toward random-state-like or random-matrix-like behavior, while variational optimization remains effective, with no evidence of exponential gradient suppression in the regime studied. We interpret this behavior in terms of locality. Spectral correlations develop at relatively shallow depth through locally generated structure, while global state randomization and the associated concentration-of-measure effects are not yet realized. These results show that commonly used statistical diagnostics of complexity do not by themselves determine trainability. Instead, they point to a separation between different aspects of complexity in finite-depth variational circuits.
\end{abstract}

\maketitle


\section{Introduction}

Variational quantum algorithms (VQAs) have become a central paradigm for near-term quantum computing because they combine parametrized quantum circuits with classical optimization and can be adapted to many applications, including ground-state preparation, quantum chemistry, combinatorial optimization, linear-algebra subroutines, and quantum simulation \cite{peruzzo2014variational,mcclean2016theory,farhi2014quantum,hadfield2019quantum,bravo2023variational,xu2021variational,mcardle2019variational,cerezo2021variational,tilly2022variational,preskill2018quantum,moll2018quantum,wecker2015progress}. In this framework, a parametrized circuit prepares a trial state $\ket{\psi(\boldsymbol{\theta})}$, and the parameters $\boldsymbol{\theta}$ are updated to minimize a cost function, often the expectation value of a Hamiltonian. For quantum many-body problems this leads to the variational quantum eigensolver (VQE), whose goal is to approximate the ground state of a target Hamiltonian.

A major obstacle is the emergence of barren plateaus, regions of parameter space in which gradients become exponentially small with increasing the number of qubits, making optimization prohibitively difficult \cite{mcclean2018barren,cerezo2021cost,holmes2022connecting,uvarov2021barren,bittel2021training}. In their simplest and most widely discussed form, barren plateaus arise when sufficiently deep or expressive parametrized circuits behave like approximate unitary $2$-designs. The severity and origin of barren plateaus depend on the cost function, circuit architecture, symmetry structure, initialization strategy, and noise model, but the central point remains that expressibility by itself does not guarantee useful trainability \cite{grant2019initialization,wang2021noise,ortiz2021entanglement,patti2021entanglement}.

This has led to structured circuit families that try to balance expressive power with optimization stability. Recent work has emphasized the role of locality and finite-depth structure in avoiding or delaying barren plateau behavior \cite{zhang2024absence,pesah2021absence,miao2024isometric}. These results raise a related question. If a structured circuit avoids a barren plateau, what kind of quantum states does it generate as the depth is increased? In particular, can trainability be inferred from statistical signatures usually associated with randomness, quantum chaos, or delocalization?

To answer this, we look beyond the variational energy. The statistical structure of a quantum state is often characterized using tools from random-matrix theory and quantum chaotic dynamics. Haar-random pure states exhibit Porter-Thomas statistics in their computational-basis probabilities (CBP)~\cite{porter1956fluctuations}, while random-matrix-like level repulsion is commonly diagnosed through adjacent-gap statistics (AGS)~\cite{oganesyan2007localization,atas2013distribution}. The inverse participation ratio (IPR) gives a complementary measure of basis localization or delocalization~\cite{santos2010onset}. These diagnostics are widely used to distinguish structured states from highly delocalized or chaotic ones \cite{shaffer2014irreversibility,iaconis2021quantum}. In many settings, their appearance is taken as evidence that a circuit is exploring Hilbert space in a nearly random way.

For the structured circuits studied here, this interpretation needs some care. We find that statistical signatures associated with random states or random-matrix behavior can appear already at shallow circuit depth through locally expressive operations. At the same time, these signatures do not necessarily imply the kind of global state randomization associated with concentration-of-measure effects. The diagnostics indicate increasing state complexity, but this increase need not coincide with a loss of useful optimization signal.

The distinction is important because different diagnostics probe different parts of the state. Probability distributions, spectral correlations, and basis spreading need not evolve at the same rate. Expressive local circuit elements can produce nontrivial correlations at shallow depth, even when the state is still far from a globally randomized state. Thus, the question is not only whether complexity appears, but how it is generated and distributed across the circuit. In this work we distinguish between locally generated complexity, arising from expressive but spatially constrained operations, and globally distributed complexity, associated with more complete Hilbert-space exploration and concentration-of-measure effects.

We examine this distinction in two many-body settings: the one-dimensional cluster-Ising model and a generalized toric code Hamiltonian on an edge-qubit lattice. For both models, we compare structured circuit families and combine conventional VQE energy benchmarks with three complementary statistical diagnostics: CBP, AGS, and IPR. Using these quantities, we track how state complexity develops with circuit depth while also monitoring the accuracy of the variational energy. The main observation is that, in the finite-depth regime studied here, increasing depth moves these diagnostics toward random-state or random-matrix benchmarks, while accurate low-energy optimization remains possible. Our work complements related studies of expressibility, gradient behavior, and optimization difficulty in variational quantum algorithms \cite{holmes2022connecting,larocca2022diagnosing,arrasmith2021effect}. 

The manuscript is organized as follows. Section II introduces the Hamiltonians and benchmark quantities. Section III describes the variational circuits, optimization workflow, and statistical diagnostics. Sections IV and V present the results for the cluster-Ising and generalized toric-code models, respectively. Section VI discusses the
relation between state complexity and trainability, including the gradient-variance diagnostics. Section VII presents the discussion and conclusions. Additional implementation details, reference values, and supporting numerical data are provided in the Appendices.


\section{Models and Benchmark Quantities}

In this section, we describe the Hamiltonians used as testbeds in this work and define the benchmark quantities used to evaluate variational performance. 

\subsection{Cluster-Ising model}

Our first testbed is the one-dimensional cluster-Ising Hamiltonian,
\begin{equation}
\HCI =
-J \sum_i Z_{i-1} X_i Z_{i+1}
-J' \sum_i Z_i Z_{i+1}
-h \sum_i X_i .
\label{eq:HCI}
\end{equation}
The first term is the cluster interaction with coupling $J$, which favors a highly entangled stabilizer-like structure associated with symmetry-protected cluster phases~\cite{raussendorf2001one}, while the second and third terms, controlled by $J'$ and $h$, introduce competing Ising and transverse-field tendencies~\cite{smacchia2011statistical}. Even in small systems, this model provides a useful setting for examining how structured ans\"atze interpolate between ordered and field-dominated regimes.

In the numerical implementation, we consider open chains with $J = J' = 1$ and a field sweep on $h$. The full dataset analyzed in this work includes chain lengths $N=8$, $12$, $14$, and $16$. In the main text we emphasize the representative sizes $N=12$ and $N=16$. Additional VQE results and statistics-only diagnostics for the other system sizes are shown in Appendices~\ref{app:cluster-vqe-additional} and \ref{app:cluster-stats-full}.

\subsection{Generalized toric code Hamiltonian}

Our second testbed is a generalized toric code Hamiltonian defined on edge qubits of a $2\times 2$ plaquette geometry. Its zero-field limit is based on the toric-code stabilizer construction introduced in Ref.~\cite{kitaev2003fault}. The physical qubits reside on the edges of the lattice, so the system contains
\begin{equation}
N = L_x (L_y + 1) + (L_x + 1) L_y = 12
\end{equation}
qubits for $L_x = L_y = 2$. The Hamiltonian is implemented as
\begin{equation}
\Htc = (1 - h)\,H_0 - \sum_{j=1}^{N}\left(h_x X_j + h_y Y_j + h_z Z_j\right),
\label{eq:HTC-general}
\end{equation}
with
\begin{equation}
H_0 = -\sum_v A_v - \sum_p B_p
\label{eq:H0-toric}
\end{equation}
and
\begin{equation}
A_v = \prod_{e \in \mathrm{star}(v)} X_e, \quad
B_p = \prod_{e \in \partial p} Z_e.
\end{equation}
Here $A_v$ is the star operator associated with vertex $v$, and $B_p$ is the plaquette operator associated with plaquette $p$. We take the field to be $h_x = h_z = h$ and $h_y = 0$, so that
\begin{equation}
\Htc = (1 - h)\left(-\sum_v A_v - \sum_p B_p\right) - h \sum_{j=1}^{N}(X_j + Z_j).
\label{eq:HTC-zx}
\end{equation}

This model provides a complementary setting to the cluster-Ising chain. In contrast to the one-dimensional system with few-body interactions, the generalized toric code Hamiltonian involves multi-qubit stabilizer operators defined on a two-dimensional lattice, leading to correlations governed by stabilizer constraints rather than short-range interactions. This choice is motivated in part by Ref.~\cite{zhang2024absence}, where
finite-local-depth circuits were benchmarked for variational training on a generalized toric code model. In the present work, we use this setting as a second testbed to ask a different question: whether the separation between state-complexity diagnostics and trainability also appears in a
stabilizer-based two-dimensional Hamiltonian.

\subsection{Exact benchmark and error measure}

For both models, exact diagonalization is used to obtain the benchmark ground-state energy $E_0$ for smaller system sizes. For larger systems ($N = 16$ in the cluster-Ising model), we use instead a Lanczos-based sparse eigensolver to access the lowest-energy eigenvalue, avoiding explicit construction of the full Hamiltonian matrix while maintaining high numerical accuracy. The variational energy is compared to $\Eexact$ both in raw form and through a normalized absolute error. To quantify variational accuracy, we define
\begin{equation}
\epsilon = \frac{|\Evqe-\Eexact|}{N}.
\label{eq:error}
\end{equation}
The normalization allows direct comparison between the two models and, for the cluster-Ising case, between different system sizes.

The use of both the raw energy per qubit and the normalized error is important. Curves of $E/N$ can appear visually close even when architectures differ meaningfully in accuracy.


\section{Variational Framework and Numerical Methods}

\subsection{Structured ansatz families}

\begin{figure*}[t]
    \centering
    \includegraphics[width=0.98\textwidth]{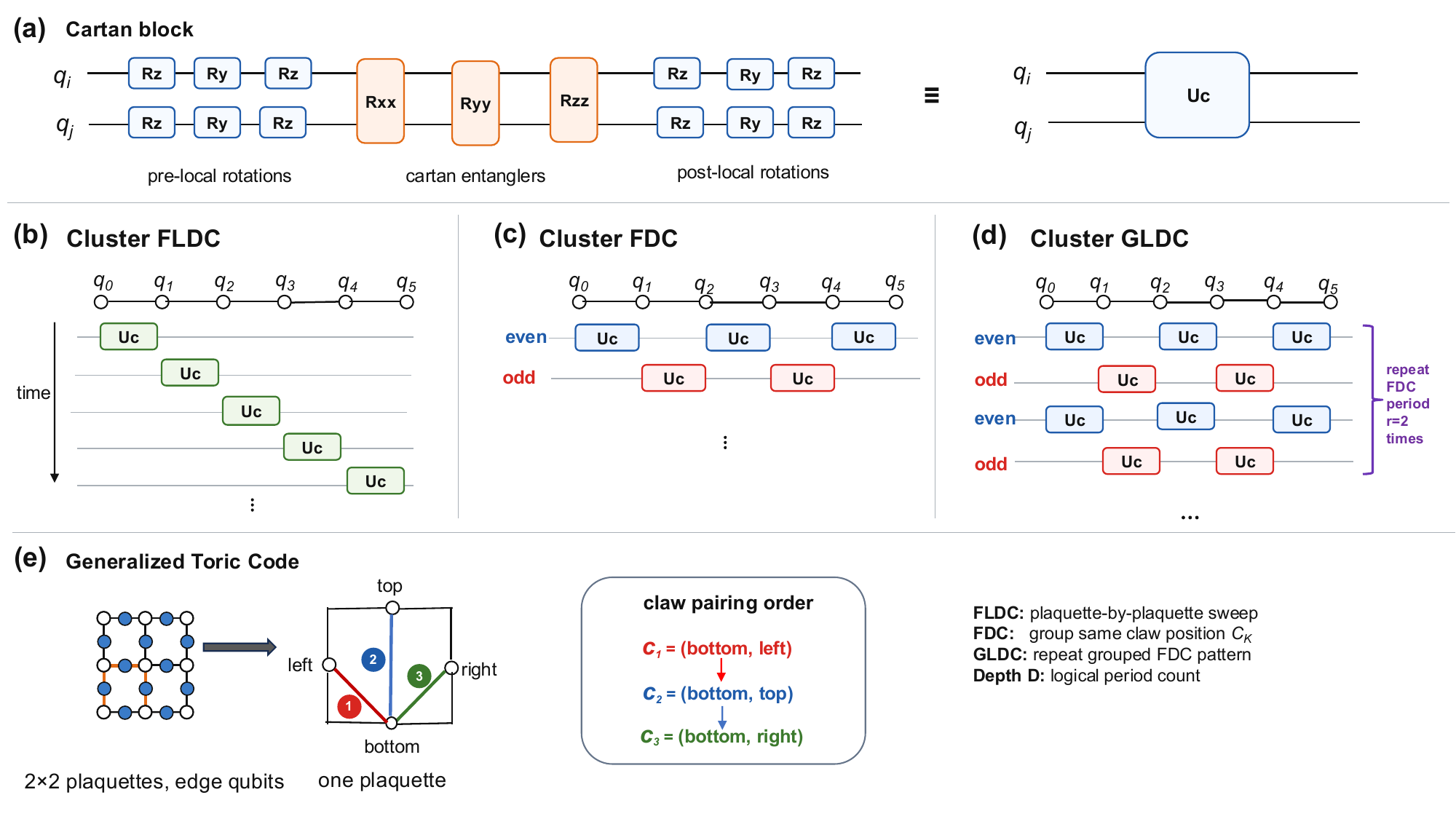}
    \caption{
    Circuit architecture used in the simulations.
    Panel (a) shows the two-qubit Cartan block, consisting of local Euler rotations, three Cartan entanglers $R_{XX}$, $R_{YY}$, and $R_{ZZ}$, and final local Euler rotations. The block is represented compactly as $U_c$ in the remaining panels. Panels (b)--(d) show the cluster-Ising scheduling: FLDC applies a sequential nearest-neighbor sweep, FDC uses grouped even/odd brick-wall layers, and GLDC repeats the FDC grouped pattern.  Panel (e) shows the generalized toric code edge-qubit layout for the $2\times2$ plaquette system with $N=12$ qubits, together with the claw ordering used within each plaquette. The depth convention used in these schedules is defined in Sec.~III.B.
    }
    \label{fig:circuit-architecture}
\end{figure*}

The calculations compare three structured ansatz families, labeled finite-depth circuits (FDC), finite-local-depth circuits (FLDC), and globally layered depth circuits (GLDC)~\cite{zhang2024absence}. The circuit schedules are summarized in Fig.~\ref{fig:circuit-architecture}. In this work, GLDC denotes a globally layered circuit obtained by repeating the FDC pattern a fixed number of times, rather than a circuit whose depth is taken to scale with system size.

All three ansatz families use the same two-qubit Cartan
block as the local variational building block, based on the
Cartan/KAK decomposition of general two-qubit operations
~\cite{zhang2003geometric,vatan2004optimal}. For qubits $q_0$ and $q_1$, the block is implemented as
\begin{align}
U_{\rm Cartan}
=&
\left[R_z(\alpha_1)R_y(\alpha_2)R_z(\alpha_3)\right]_{q_0} 
\nonumber\\
&\otimes
\left[R_z(\beta_1)R_y(\beta_2)R_z(\beta_3)\right]_{q_1}
\nonumber\\
&\times
R_{XX}(\theta_x)R_{YY}(\theta_y)R_{ZZ}(\theta_z)
\nonumber\\
&\times
\left[R_z(\alpha_4)R_y(\alpha_5)R_z(\alpha_6)\right]_{q_0}
\nonumber\\
&\otimes
\left[R_z(\beta_4)R_y(\beta_5)R_z(\beta_6)\right]_{q_1}.
\end{align}
Here
\begin{equation}
R_\alpha(\theta)=\exp\left(-i\frac{\theta}{2}\alpha\right),
\qquad
\alpha\in\{X,Y,Z\},
\end{equation}
and
\begin{equation}
R_{\alpha\alpha}(\theta)=
\exp\left(-i\frac{\theta}{2}\alpha\otimes\alpha\right).
\end{equation}
The three entangling angles $\{\theta_x,\theta_y,\theta_z\}$ and the twelve single-qubit Euler angles give 15 variational parameters per block.

The Cartan block is already locally expressive: a single two-qubit block can create entanglement and nontrivial local correlations. Thus, even shallow circuits made from these blocks can generate states with local structure. The distinction among FDC, FLDC, and GLDC lies in how these two-qubit blocks are scheduled across the system. For the cluster-Ising chain, FLDC applies nearest-neighbor Cartan blocks in a sequential sweep along the chain. By contrast, FDC uses grouped even- and odd-bond brick-wall layers, and GLDC repeats this grouped FDC pattern.

For the generalized toric code lattice, the plaquette geometry is used to organize the two-qubit gates. In the claw ordering used here, the three pairings associated with a plaquette are applied in the sequence shown in panel~(e) of Fig.~\ref{fig:circuit-architecture}. FLDC applies these blocks plaquette by plaquette, FDC groups blocks according to their position within each plaquette, and GLDC repeats the grouped FDC pattern. Since same-position groups are not always pairwise disjoint on the open $2\times2$ geometry, the implementation decomposes each such group into the minimal number of disjoint sublayers required by qubit overlap. This keeps the intended grouped structure while producing a valid gate schedule.

These scheduling choices matter because trainability is not controlled only by the number of parameters. It also depends on how the circuit spreads correlations and explores Hilbert space. A globally grouped construction can generate random-state-like features more rapidly, whereas a finite-local-depth sweep can preserve more geometrically organized structure over the same nominal depth range.

\subsection{Depth, parameter count, and logical repetitions}

The notion of depth is implemented at the level of a logical period of the requested ansatz. For FLDC, one period corresponds to a single sequential sweep of the relevant two-qubit blocks; for FDC, one period corresponds to one pass through the grouped layers. GLDC is constructed by repeating the FDC pattern, as illustrated in Fig.~\ref{fig:circuit-architecture}. 

In the simulations reported here, GLDC is implemented using explicit repetition counts, typically $r=2$ or $r=3$. With this convention, $r=1$ is identical to the corresponding FDC circuit, while larger $r$ values repeat the same grouped FDC pattern. This allows for a controlled comparison between circuit families at comparable effective depth, rather than relying on system-size-dependent repetition.

If $n_{\mathrm{blk}}$ denotes the number of two-qubit Cartan blocks in one logical period and $r_{\mathrm{eff}}$ denotes the effective number of period repetitions, then the VQE parameter count is
\begin{equation}
N_{\theta}=15\,n_{\mathrm{blk}}\,r_{\mathrm{eff}}.
\label{eq:param-count}
\end{equation}

The same scheduling logic is used in the statistics-only calculations described below, where the parameters of the Cartan blocks are sampled randomly to generate ensembles of circuit outputs. In this way, the statistical diagnostics probe intrinsic circuit-induced complexity independently of the variational optimization procedure.

\subsection{VQE workflow and optimization}

For a parametrized state $\ket{\psi(\boldsymbol{\theta})}$, the VQE objective is
\begin{equation}
E(\boldsymbol{\theta})=\bra{\psi(\boldsymbol{\theta})}H\ket{\psi(\boldsymbol{\theta})},
\label{eq:cost}
\end{equation}
where $H$ is either the cluster-Ising Hamiltonian or the generalized toric code Hamiltonian. The motivating barren plateau picture is that for sufficiently expressive circuits, the variance of a gradient component may decay with the number of qubits $N$ as
\begin{equation}
\mathrm{Var}\!\left(\frac{\partial E}{\partial \theta_k}\right)\sim \order{2^{-N}},
\label{eq:bp}
\end{equation}
or, more generally, become exponentially small in system size under suitable assumptions \cite{mcclean2018barren,cerezo2021cost,holmes2022connecting}. Equation~\eqref{eq:bp} does not by itself determine trainability in every finite problem, but it motivates the search for observables that might warn of an impending loss of optimization signal.

We evaluate the VQE cost using TensorCircuit and JAX~\cite{zhang2023tensorcircuit,bradbury2018jax}. Rather than constructing dense Hamiltonian matrices during optimization, we encode the Hamiltonian as a sparse list of Pauli strings. The energy is then computed as a sum of Pauli-string expectation values,
\begin{equation}
E(\boldsymbol{\theta})=\sum_t c_t \,
\bra{\psi(\boldsymbol{\theta})} P_t \ket{\psi(\boldsymbol{\theta})},
\label{eq:sparse-energy}
\end{equation}
where each $P_t$ acts nontrivially only on the support of the corresponding Hamiltonian term. This matrix-free evaluation is particularly natural for the present models, where the interaction terms have bounded locality.

We obtain gradients by automatic differentiation, meaning that the derivatives of the energy are evaluated by differentiating the computational graph used for the circuit and expectation-value calculation. Each independent training trajectory is then optimized with Adam~\cite{kingma2014adam}. In the numerical implementation, we use the standard Adam moments $(m_t,v_t)$ with learning rate $\eta=10^{-2}$ and coefficients $(\beta_1,\beta_2)=(0.9,0.999)$. Parameters are initialized uniformly in
$[0,2\pi)$, and multiple random restarts are performed for each value of the field strength. The VQE calculations use $100$ random restarts for each value of the field strength, and each restart is run for $1000$ Adam iterations. For each field value, we report the best variational energy, together with summary statistics over the optimized restarts. The resulting energy $\Evqe$ is then compared with the benchmark ground-state energy $\Eexact$.

\subsection{State-complexity diagnostics}

The optimized energy provides an important benchmark for accuracy, yet it does not fully characterize the structure of the variational state. Two ans\"atze may achieve similar energies while producing states with very different statistical properties.

This motivates the use of diagnostics that probe different aspects of complexity:
\begin{enumerate}
    \item \textbf{Amplitude statistics}, through the computational-basis probability distribution 
    ~\cite{porter1956fluctuations,hayden2006aspects};

    \item \textbf{Spectral correlations}, through the entanglement spectrum 
~\cite{li2008entanglement,oganesyan2007localization,atas2013distribution};

    \item \textbf{Basis delocalization}, through the inverse participation ratio
    ~\cite{santos2010onset}.
\end{enumerate}
Each diagnostic reflects a different notion of randomness and complements direct measures of trainability, such as gradient-variance and optimization-landscape analyses
~\cite{larocca2022diagnosing,arrasmith2021effect}. Agreement between them suggests increasing complexity, whereas disagreement can also be informative.

\subsection{Porter-Thomas statistics}

For a pure state expanded in the computational basis,
\begin{equation}
\ket{\psi}=\sum_{z=0}^{2^N-1} c_z \ket{z},
\end{equation}
the basis probabilities are $p_z=|c_z|^2$. For Haar-random states, and in the large-$D$ limit up to the normalization constraint, the amplitudes $c_z$ in any fixed basis behave as normalized complex Gaussian variables. As a result, the rescaled probabilities $y_z=D\, p_z$, with $D=2^N$, follow the unit-rate exponential distribution,
\begin{equation}
P(y_z)=e^{-y_z}.
\end{equation}
Equivalently, the probability density for $p_z$ takes the Porter-Thomas form
\cite{porter1956fluctuations},
\begin{equation}
P(p_z)=D\, e^{-Dp_z}.
\label{eq:PT}
\end{equation}
This form is widely used as a benchmark in random-circuit sampling and cross-entropy benchmarking~\cite{arute2019quantum}. In the numerical workflow, the comparison to Porter-Thomas statistics is made through a Kolmogorov-Smirnov (KS) distance between the empirical distribution of $y_z$ and the unit-rate exponential distribution. A decrease in this distance indicates that the computational-basis amplitudes are becoming more random-state-like.

This diagnostic probes whether basis probabilities are broadly distributed as expected for random states. It is sensitive to amplitude statistics but does not by itself reveal spectral correlations or entanglement structure.

\subsection{Entanglement-spectrum adjacent-gap ratio}

To probe the internal structure of entanglement, we bipartition the system into subsystems $A$ and $B$ and compute the reduced density matrix
\begin{equation}
\rho_A = \Tr_B \left( \ket{\psi}\bra{\psi} \right).
\end{equation}
In the present calculations, the bipartition is chosen as an equal cut with $N_A=\lfloor N/2 \rfloor$ qubits. Let $\{\lambda_n\}$ denote the nonzero eigenvalues of $\rho_A$, ordered decreasingly. Following the entanglement-spectrum convention employed in Ref.~\cite{li2008entanglement}, we define the corresponding entanglement energies as 
\begin{equation}
\xi_n=-\log \lambda_n .
\end{equation}
From adjacent spacings $\delta_n=\xi_{n+1}-\xi_n$, one constructs the adjacent-gap-ratio statistic~\cite{oganesyan2007localization,atas2013distribution}, 
\begin{equation}
r_n = \frac{\min(\delta_n,\delta_{n-1})}{\max(\delta_n,\delta_{n-1})}.
\label{eq:rstat}
\end{equation}
The mean $r_n$ distinguishes between uncorrelated and level-repelling spectra: Poisson-like statistics give $\langle r_n\rangle \simeq 0.386$, while GOE and GUE level statistics give $\langle r_n\rangle \simeq 0.536$ and $\langle r_n\rangle \simeq 0.603$, respectively
~\cite{atas2013distribution}. 

In the present context, the $r_n$ ratio is useful here because it probes spectral organization rather than only basis probabilities. When $r_n$ moves toward random-matrix-like values while the VQE energy remains accurate, it suggests that complexity diagnostics and trainability need not track one another in a simple way. For the generic complex Cartan-block circuits studied here, we use the GUE value as the level-repulsion reference and the Poisson value as the uncorrelated-spectrum reference in the figures below.

\subsection{Inverse participation ratio}

The inverse participation ratio in the computational basis is a standard basis-localization diagnostic in studies of quantum chaos and thermalization~\cite{santos2010onset}, and is defined here as
\begin{equation}
\mathrm{IPR}=D\sum_z |c_z|^4,
\qquad D=2^N .
\label{eq:IPR}
\end{equation}
Notice that $1\leq {\rm IPR}\leq D$. Larger IPR indicates stronger localization in the chosen basis, while smaller IPR indicates delocalization over many basis states. With the normalization used here, a perfectly uniform equal-amplitude state has ${\rm IPR}=1$, whereas a Haar-random complex state has $\langle {\rm IPR}\rangle\to 2$ in the large Hilbert-space limit; see Appendix~\ref{app:ipr}. Thus, values near $2$ should be interpreted as random-state-like amplitude statistics, not as a lack of delocalization.

The IPR complements Porter-Thomas statistics. Both concern basis amplitudes, but the IPR compresses that information into a single localization measure, whereas the Porter-Thomas comparison is sensitive to the overall shape of the probability distribution.

While these diagnostics are often used as indicators of complexity or randomness, they probe distinct aspects of quantum-state structure and need not evolve in a strictly correlated manner. In particular, agreement among multiple diagnostics strengthens the interpretation of increased complexity, whereas discrepancies between them can signal intermediate regimes in which different notions of complexity emerge at different rates.

\subsection{Random-parameter circuit ensembles}

To separate architecture-induced state complexity from the optimization procedure, we also study ensembles of circuits whose parameters are sampled independently and uniformly from $[0,2\pi)$. These ensembles use the same Cartan blocks and gate schedules as the VQE ans\"atze. The depth-dependent
Porter-Thomas distance, entanglement-spectrum statistic, and IPR reported below are computed from these random-parameter ensembles. They therefore characterize the statistical structure generated by the circuit architecture,
rather than the detailed distribution of states reached after optimization.


\section{Results for the Cluster-Ising Model}

\subsection{Variational energy performance}

\begin{figure}[ht]
\centering
\includegraphics[width=0.96\columnwidth]{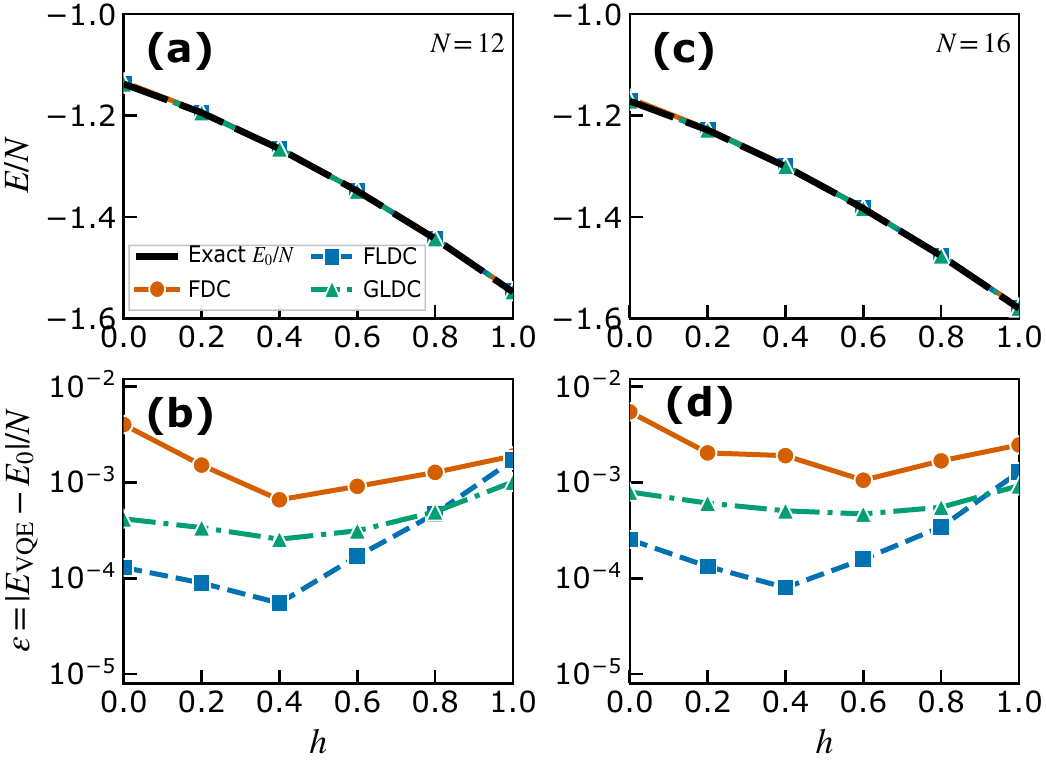}
\caption{
Cluster-Ising VQE results for $N=12$ [panels (a) and (b)] and $N=16$ [panels (c) and (d)]. The upper panels show the optimized energy per qubit, while the lower panels show the normalized error $\epsilon=|E_{\rm VQE}-E_0|/N$. The FDC and FLDC curves use one logical period, whereas the GLDC curve uses two repetitions of the grouped FDC pattern, $r=2$. All VQE calculations use 100 random restarts and 1000 Adam iterations per restart.
}
\label{fig:CI_VQE}
\end{figure}

Figure~\ref{fig:CI_VQE} summarizes the cluster-Ising VQE performance for representative system sizes $N=12$ and $N=16$ across the field sweep. For both system sizes, the optimized energies obtained from the structured circuit families lie close to the exact benchmark on the scale of the $E/N$ plots. The normalized error gives a more sensitive view of the same data. It shows that the ansatz choice still matters, even in field regions where the raw energy curves almost overlap. We therefore report both quantities in the comparisons below.

\begin{figure*}[ht]
    \centering
    \includegraphics[width=0.9\textwidth]{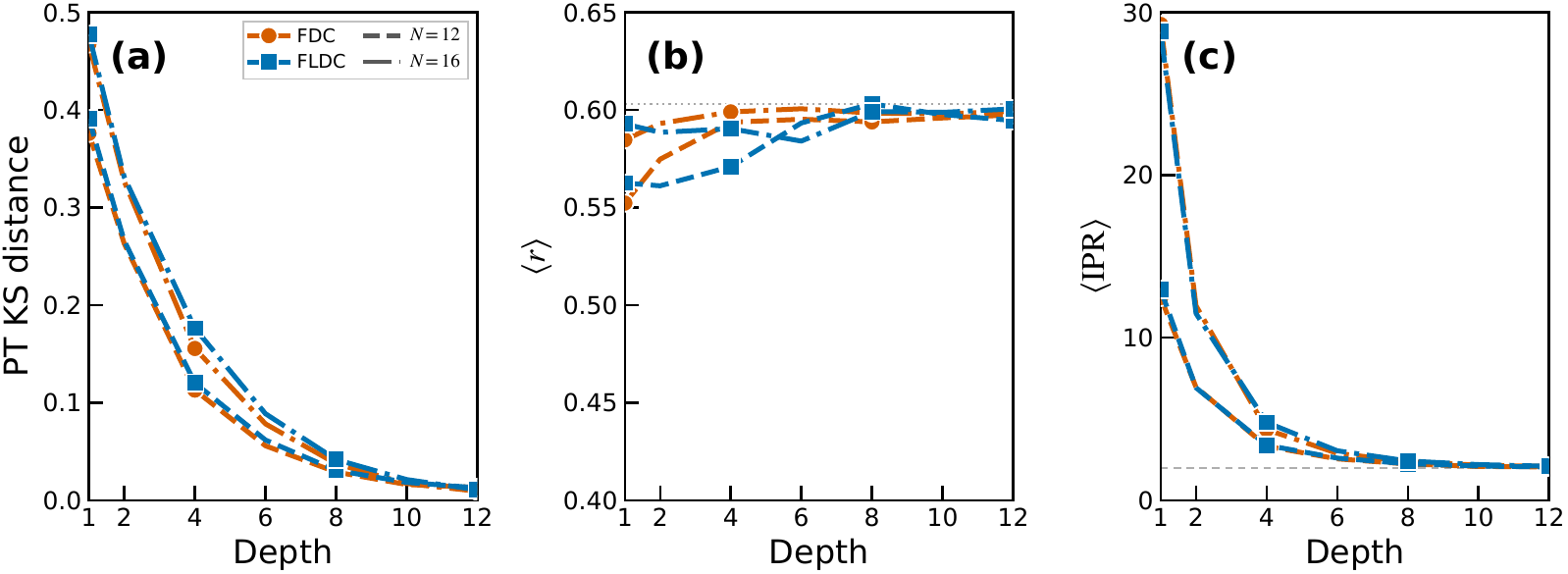}
    \caption{
Cluster-Ising statistics-only diagnostics for random-parameter FDC and FLDC ensembles at $N=12$ and $N=16$. The panels show: (a) the Porter-Thomas KS distance, (b) the entanglement-spectrum adjacent-gap average ratio $\langle r\rangle$, and (c) the normalized average inverse participation ratio $\langle{\rm IPR}\rangle$. In panel (b), the dotted horizontal line marks the GUE reference value. In panel (c), the horizontal reference line marks the random-state-like delocalization value $\langle{\rm IPR}\rangle\simeq 2$ for the normalization used here. Each point is averaged over 100 random circuit samples.
}
    \label{fig:CI_stats}
\end{figure*}

\subsection{Depth dependence of the statistical diagnostics}

Figure~\ref{fig:CI_stats} shows the statistical diagnostics as a function of circuit depth for the cluster-Ising model at $N=12$ and $N=16$. These diagnostics are evaluated on randomly parameterized circuit ensembles built from the same ansatz families, rather than on the optimized VQE states. They therefore probe the statistical structure generated by the circuit architecture itself. GLDC is not included in Fig.~\ref{fig:CI_stats}. Since GLDC is constructed by repeating the grouped FDC pattern, the main statistics-only comparison here is between the grouped FDC schedule and the sequential FLDC schedule.

The Porter-Thomas KS distance decreases with depth for both system sizes, showing that the computational-basis probability distribution moves closer to the exponential form expected for random pure states. The entanglement-spectrum average ratio also increases toward values associated with level repulsion. This change happens rather quickly, already at relatively shallow depth. The IPR decreases with depth as well, approaching the random-state-like delocalization value $\langle{\rm IPR}\rangle\simeq 2$ for the normalization used here, indicating that the sampled states spread over more computational-basis configurations.

The three diagnostics do not move at exactly the same rate. The entanglement-spectrum statistic is the fastest to approach a random-matrix-like regime, while the Porter-Thomas KS distance and the IPR retain visible finite-depth structure over the same range. This separation indicates that spectral correlations, amplitude statistics, and basis delocalization are not identical measures of circuit-induced complexity.

\subsection{System-size comparison: representative sizes and supplementary data}

The main text focuses on $N=12$ and $N=16$. Additional VQE results for $N=8$ and $N=14$ are shown in Appendix~\ref{app:cluster-vqe-additional}, and the full statistics-only data are collected in Appendix~\ref{app:cluster-stats-full}. Across these sizes, the qualitative behavior is consistent. The VQE energies remain close to the exact benchmarks, and the statistics-only ensembles show the same depth-dependent trends: the Porter-Thomas KS distance decreases toward zero, the entanglement-spectrum average ratio moves toward universal level-repelling values, and the IPR is reduced toward its Haar measure value. These results should be read as a finite-size comparison rather than an asymptotic scaling analysis. Larger systems would be needed to determine the eventual large-$N$ behavior. 

\subsection{Interpretation for the cluster-Ising data}

For the cluster-Ising model, the same ansatz families that give accurate low-energy VQE results also generate stronger statistical signatures of complexity when the random-parameter circuit depth is increased. One should not read this as evidence that the sampled states have become Haar-random, or that the circuits form approximate unitary designs. The data support a narrower conclusion. In these structured finite-depth circuits, spectral correlations and basis spreading can appear clearly before the circuit reaches a regime dominated by global concentration-of-measure effects. Thus, the movement of the Porter-Thomas distance, entanglement-spectrum statistic, and IPR toward their reference values is not, by itself, a sign that the optimization landscape has become barren.


\section{Results for the Generalized Toric Code Hamiltonian}

\subsection{Variational energy performance}

\begin{figure}[ht]
    \centering
    \includegraphics[width=0.85\columnwidth]{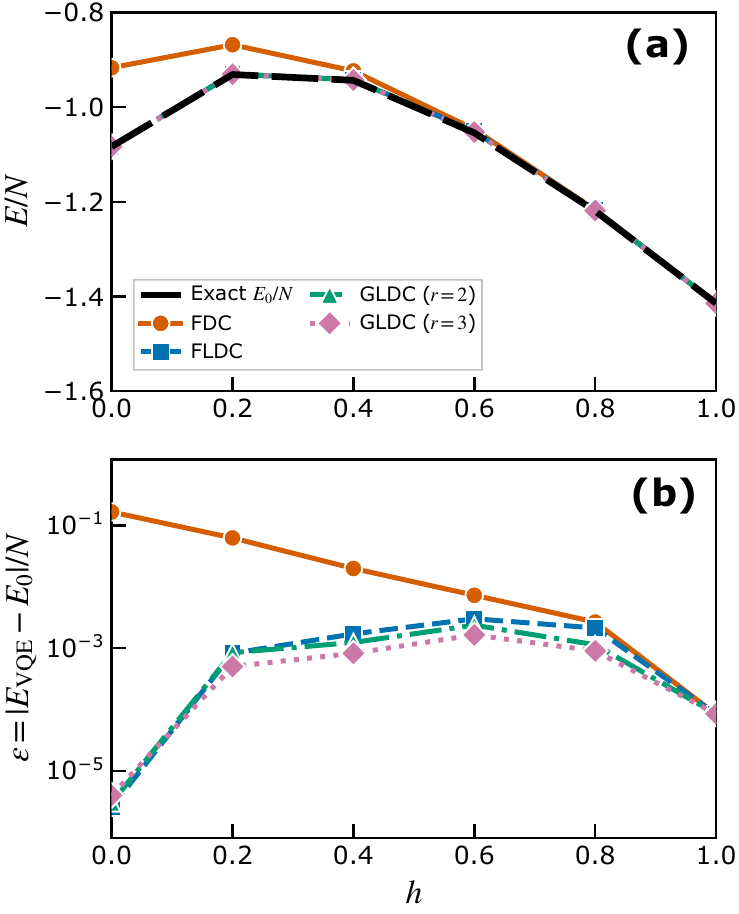}
    
    \caption{Generalized toric-code VQE results for the $N=12$ edge-qubit geometry with claw within-plaquette ordering. Panel (a) shows the optimized energy per qubit, and panel (b) shows the normalized error
    $\epsilon=|E_{\rm VQE}-E_0|/N$. The FDC and FLDC curves use one logical period, whereas the GLDC curves use two and three repetitions of the grouped FDC pattern, $r=2$ and $r=3$, respectively. All VQE calculations use $100$ random restarts and $1000$ Adam iterations per restart.}

    \label{fig:TC_VQE}
\end{figure}

The generalized toric code Hamiltonian gives a second test case with a different operator structure and geometry. Figure~\ref{fig:TC_VQE} shows the energy per qubit and normalized error for the ansatz families studied at $N=12$.

The optimized energies follow the exact benchmark closely over the range of fields considered. As in the cluster-Ising case, the normalized error is more informative than the raw energy alone. The differences between ansatz families are most visible at smaller field values, where the stabilizer structure inherited from the toric-code limit is still important. The repeated grouped circuits are useful in this regime. Compared with the FDC circuit, the GLDC variants with $r=2$ and $r=3$ substantially reduce the normalized error at low fields and remain close to the best curves. This indicates that a small number of structured grouped repetitions can improve the variational accuracy without replacing the ansatz by an unrestricted or fully random circuit.

This model is not meant to provide an additional system-size scaling study, since the geometry used here fixes the number of edge qubits. Its role is instead to test whether the same qualitative behavior appears in a stabilizer-based Hamiltonian with a different microscopic structure from the one-dimensional cluster-Ising chain.

\begin{figure*}[ht]
    \centering
    \includegraphics[width=0.9\textwidth]{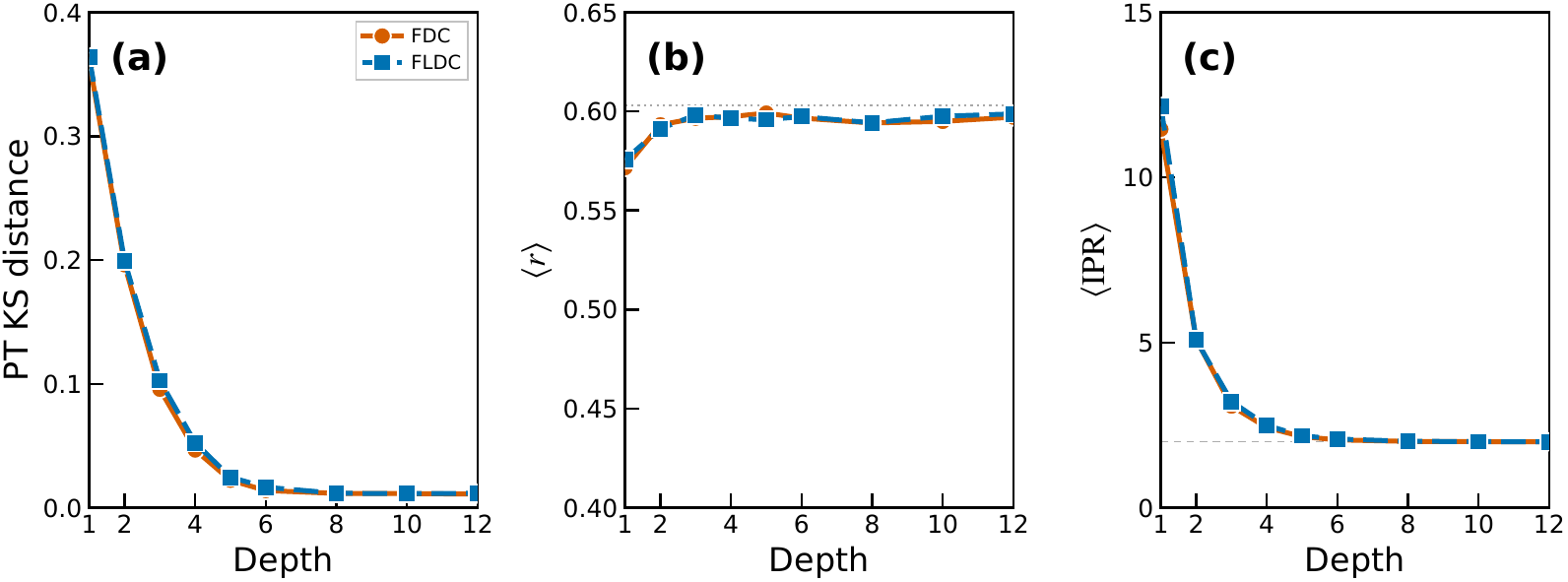}
    \caption{
    Statistics-only diagnostics for the generalized toric-code ansatz with claw within-plaquette ordering. The panels show: (a) the Porter-Thomas KS distance, (b) the entanglement-spectrum adjacent-gap ratio $\langle r\rangle$, and (c) the normalized inverse participation ratio $\langle{\rm IPR}\rangle$. In panel (b), the dotted horizontal line marks the GUE reference value. In panel (c), the horizontal reference line marks the random-state-like delocalization value $\langle{\rm IPR}\rangle\simeq 2$ for the normalization used here. Each point is averaged over 200 random circuit samples.}
    \label{fig:toric_stats}
\end{figure*}

\subsection{Depth dependence of the statistical diagnostics}

Figure~\ref{fig:toric_stats} displays the Porter-Thomas KS distance, the entanglement-spectrum average ratio, and the IPR for the generalized toric code circuit ensembles as a function of depth.

The trends are similar to those found for the cluster-Ising circuits. The Porter-Thomas KS distance decreases with depth, the entanglement-spectrum average ratio moves toward universal level-repelling values, and the IPR decreases toward the same random-state reference discussed above. Thus, the randomly sampled states generated by the toric-code ansatz become more delocalized in the computational basis and develop stronger spectral correlations as the circuit depth is increased.

We also note that the Porter-Thomas KS distance seems to level off at a small but nonzero value over the depths studied. This suggests that the finite-depth toric-code circuits move toward Porter-Thomas statistics but do not fully reach them in this finite-size regime. While we do not have a conclusive interpretation for it, it is possible that the relatively small lattice size and limited number of samples prevent the distribution from fully reaching the Porter-Thomas functional form for this particular toric-code circuit ensemble.



\section{State Complexity Versus Trainability}

The preceding sections compare two related but distinct aspects of the circuits. The energy and gradient data probe optimization, while the Porter-Thomas distance, entanglement-spectrum statistic, and IPR probe the statistical structure of states generated by the ansatz. These quantities relate to each other, but are not equivalent. A circuit may generate states with nontrivial statistical features without immediately producing an optimization landscape with vanishing gradients.

\begin{figure}[ht]
    \centering
    \includegraphics[width=\columnwidth]{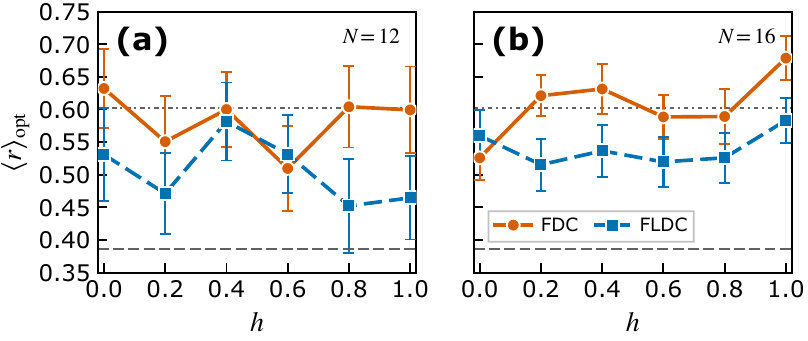}
    \caption{
Entanglement-spectrum adjacent-gap ratio of the optimized cluster-Ising VQE states at depth $D=1$. Panels (a) and (b) show $\langle r\rangle_{\rm opt}$ as a function of field strength for $N=12$ and $N=16$, respectively. Error bars denote the standard error of the mean adjacent-gap ratio computed from the entanglement spectrum of each optimized state. The dotted and dashed horizontal lines mark the GUE and Poisson reference values, respectively.
}

    \label{fig:cluster_optstate_r}
\end{figure}

As a direct check on the trained states, we also compute the entanglement-spectrum adjacent-gap ratio for the optimized VQE state at each field value. These are the same depth $D=1$ cluster-Ising VQE runs shown in Fig.~\ref{fig:CI_VQE}; the $N=12$ benchmarks use exact diagonalization, while the $N=16$ benchmarks use the Lanczos sparse eigensolver described in Sec.~II.C. Figure~\ref{fig:cluster_optstate_r} shows that the optimized states already have level-repelling values of $\langle r\rangle_{\rm opt}$, generally lying well above the Poisson reference and close to the GUE value across the field sweep. The error bars are comparable to the small oscillations of the average values with field strength, indicating that these fluctuations should not be overinterpreted. Thus, the low-energy states found by the optimizer carry nontrivial entanglement-spectrum correlations, even though the gradient-variance data below do not show the rapid system-size suppression expected from a barren plateau in the studied finite-size regime.

Locality gives a natural way to understand why this can happen. In a finite-depth geometrically local circuit, the derivative with respect to a local gate parameter is affected only by Hamiltonian terms that overlap the corresponding causal light cone. At fixed depth, this region remains small compared with the full system. This is the same locality-based intuition that appears in analyses of local-cost barren plateaus and finite-local-depth trainability~\cite{cerezo2021cost,uvarov2021barren,zhang2024absence}. Gradient statistics are therefore governed by finite subsystems rather than by concentration over the full Hilbert space. The same finite-depth structure also limits the operator directions that are accessible to the circuit. In the language of Lie-algebraic analyses of trainability, the relevant accessible directions at fixed depth form only a restricted part of the full many-body operator space~\cite{ragone2024lie}. This argument is used to interpret the finite-size numerical results, not as an independent proof of asymptotic trainability; see Appendix~\ref{app:lightcone-accessibility}.

To probe this trainability aspect directly, we compute gradient statistics for the cluster-Ising model. For each system size and depth, we evaluate the variance of each energy-gradient component over 100 randomly initialized circuit-parameter samples and then average over all trainable parameters,
\begin{equation}
\overline{\mathrm{Var}}(\partial_\theta E)
=
\frac{1}{N_\theta}
\sum_{k=1}^{N_\theta}
\mathrm{Var}_\theta
\left(
\frac{\partial E}{\partial \theta_k}
\right).
\end{equation}


\begin{figure}[t]
\centering
\includegraphics[width=\columnwidth]{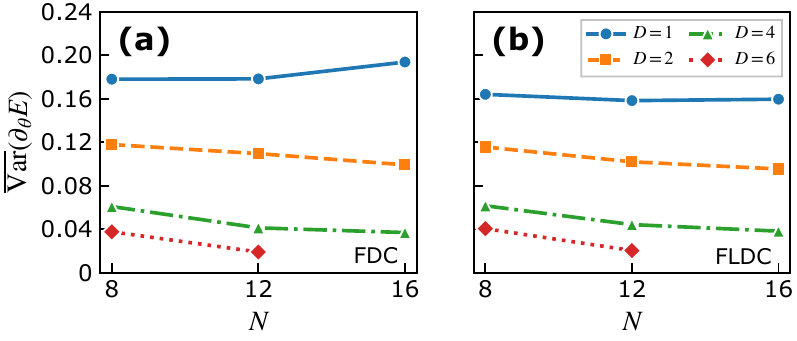}
\caption{
Mean per-parameter gradient variance for the cluster-Ising model at field strength $h=0.6$, with $J=J'=1$. Panels (a) and (b) show FDC and FLDC, respectively, with different curves corresponding to circuit depths $D=1,2,4$, and $6$. Each point is estimated from 100 random parameter samples.
}
\label{fig:grad_var_vs_N}
\end{figure}

An exponential decrease of this quantity with system size would be a signature of barren-plateau-like gradient suppression. Figure~\ref{fig:grad_var_vs_N} shows mean per-parameter gradient variance as a function of system size for several fixed circuit depths. For both FDC and FLDC, over the available sizes $N=8,12,16$, we do not observe a rapid collapse of the variance with $N$. The more visible trend is the decrease of the typical per-parameter variance with increasing depth, as expected for more expressive circuits. Within this finite-size window, the movement of the state diagnostics toward random-state or random-matrix benchmarks is therefore not accompanied by the gradient suppression expected from a barren plateau.

The diagnostics also illustrate why a single notion of ``complexity'' is too coarse. The entanglement-spectrum statistic reaches random-matrix-like values rather quickly, which is consistent with the strong local expressivity of the Cartan blocks. Similar behavior has been observed in many-body settings where entanglement spectra display random-matrix correlations even though the states retain nontrivial structure~\cite{chamon2014emergent,yang2015two,yang2017entanglement,santos2010onset}. The amplitude-based diagnostics evolve more gradually. In particular, the Porter-Thomas distance and the IPR show that the computational-basis probabilities are still changing over a depth range where the entanglement-spectrum correlations have already become close to their random-matrix benchmark.

\FloatBarrier
\section{Discussion and Conclusions}

The results separate the appearance of random-state-like diagnostics from barren-plateau behavior. In the finite-depth circuits studied here, the statistical diagnostics move toward their random-state or random-matrix reference values, while the VQE performance remains accurate and the gradient data do not show clear exponential suppression over the available system sizes.

This suggests a more careful view of the usual connection between expressivity and trainability in variational quantum algorithms. Expressivity is not simply good or bad for optimization. What matters is how the circuit generates and spreads correlations. For optimization problems, the implication for ansatz design is therefore not simply to suppress expressivity, but to organize it through locality, depth, and gate scheduling. Circuits can be locally rich enough to represent nontrivial many-body states while still avoiding, at least over a finite-depth window, the kind of global scrambling that typically leads to barren-plateau behavior. The finite-local-depth result of Ref.~\cite{zhang2024absence} is especially close to this viewpoint, since it shows that circuits may remain trainable even while being capable of generating nontrivial long-range entanglement. Our results are consistent with that picture, but they address a complementary question: the same circuits can already show statistical signatures usually associated with complex or random states before the onset of global concentration effects.

Recent dynamic-circuit approaches provide another useful comparison. Dynamic parameterized quantum circuits, which include intermediate measurements and feedforward operations, have been shown to combine strong expressivity with provable absence of barren plateaus~\cite{deshpande2024dynamic}. 
Work on dynamic-circuit characterization and benchmarking, including Ref.~\cite{shirgure2026characterizing}, also highlights that mid-circuit measurements and feedforward change the circuit structure and hardware tradeoffs relative to static unitary circuits. This is not in conflict with our interpretation; it is simply a different mechanism. The circuit model itself is changed by including measurement and feedforward, whereas the present work studies static structured ans\"atze and asks how their state-complexity diagnostics relate to trainability. 
In this sense, dynamic-circuit constructions and finite-local-depth static circuits both show that expressivity and barren plateaus are not tied together in a simple one-to-one way, but they separate them by different physical mechanisms.

There are also limits to this conclusion. The calculations are restricted to finite systems and accessible circuit depths. At larger sizes, or at much greater depths, the same circuit families may eventually enter a regime where global concentration effects become important. The present results therefore do not imply that complexity and barren plateaus are unrelated. They show instead that the relationship depends on how the complexity is produced.

Several directions remain open. Larger systems would give a better test of scaling, and a more systematic study of gradient variance as a function of both depth and system size would help identify the crossover from local complexity to global concentration. It would also be useful to compare different circuit geometries, initialization schemes, and training strategies~\cite{grant2019initialization,sung2020using,volkoff2021large,liu2024mitigating} to determine how robust this separation is beyond the ansatz families considered here.



\begin{acknowledgments}
We thank J. Reyes for support during the early stages of this work.
\end{acknowledgments}


\appendix

\section{Implementation details relevant for reproducibility}

We summarize several implementation details that are used in the numerical workflow.

For the cluster-Ising calculations, the chain geometry is represented by nearest-neighbor bonds, with FLDC implemented as a sequential sweep over those bonds and FDC/GLDC implemented through even-bond and odd-bond brick-wall groups. For the generalized toric code calculations on the $2\times 2$ plaquette geometry, the qubits reside on edges, and the plaquette-level scheduling is organized by a fixed within-plaquette ordering. 

The VQE optimization uses JAX-based automatic differentiation and a jit-compiled Adam loop~\cite{bradbury2018jax}. Hamiltonian expectation values are evaluated from sparse Pauli-string encodings rather than by repeated construction of dense matrices~\cite{cerezo2021variational,tilly2022variational}. For each field value, exact diagonalization results are cached and reused across ansatz runs, ensuring that all variational comparisons are made against the same exact benchmark. 

The implementation also allows for additional quantities relevant to benchmarking and validation, including the fidelity with the exact ground state and consistency checks between sparse and dense energy evaluations. These features provide standard validation of the numerical workflow and are consistent with best practices in VQE implementations~\cite{cerezo2021variational,tilly2022variational}. While such checks are useful for numerical reliability, the main physical conclusions of this work are already captured by the variational energy, normalized error, and the statistical diagnostics discussed in the main text.


\section{Light-cone structure and finite-depth accessibility}
\label{app:lightcone-accessibility}

This appendix shows the locality argument used in Sec.~VI. For a local Hamiltonian and a geometrically local circuit with bounded local depth, a gradient with respect to a local parameter is controlled by the part of the Hamiltonian lying in the corresponding causal region of the circuit. Related locality arguments appear in Refs.~\cite{cerezo2021cost,uvarov2021barren,zhang2024absence}. We also state the corresponding restriction on the operator directions accessible at fixed depth, following the Lie-algebraic viewpoint of Ref.~\cite{ragone2024lie}. In this appendix, $D$ refers to the local depth relevant for the light-cone argument, namely the number of overlapping local layers through which operator support can propagate. This should be distinguished from the logical period count used to label the numerical circuit schedules in Sec.~III.B; the argument below is used as a locality interpretation of the finite-size results, not as an independent proof of asymptotic trainability for every schedule.

Consider a local cost function
\begin{equation}
C(\boldsymbol{\theta})
=
\langle \psi(\boldsymbol{\theta})|
H_{\rm loc}
|\psi(\boldsymbol{\theta})\rangle,
\qquad
H_{\rm loc}
=
\sum_{\ell} h_{\ell},
\end{equation}
where each $h_{\ell}$ has bounded support independent of the total system size $N$.

\noindent\textit{Proposition 1. Local-gradient light cone.}
Let the parameter $\theta_a$ appear in a local gate $\exp(-i\theta_a G_a)$, where $G_a$ has bounded support. For a geometrically local circuit with fixed local depth $D$, the gradient $\partial C/\partial\theta_a$ receives contributions only from Hamiltonian terms lying in the causal neighborhood of that gate. The size of this neighborhood is independent of $N$.

\noindent\textit{Proof.}
Write the circuit as
\begin{equation}
U(\boldsymbol{\theta})
=
U_{>a}
e^{-i\theta_a G_a}
U_{<a},
\end{equation}
where $U_{<a}$ and $U_{>a}$ contain the gates before and after the parameterized gate. Then
\begin{equation}
\frac{\partial C}{\partial \theta_a}
=
i
\langle 0|
U_{<a}^{\dagger}
e^{i\theta_a G_a}
\left[
G_a,
U_{>a}^{\dagger}H_{\rm loc}U_{>a}
\right]
e^{-i\theta_a G_a}
U_{<a}
|0\rangle .
\end{equation}
Since $H_{\rm loc}=\sum_{\ell}h_{\ell}$, a term $h_{\ell}$ contributes only if the support of
$U_{>a}^{\dagger}h_{\ell}U_{>a}$
overlaps the support of $G_a$. Otherwise the commutator vanishes.

Geometric locality and fixed local depth imply that
$U_{>a}^{\dagger}h_{\ell}U_{>a}$
can spread only within a depth $D$ causal region. Therefore, only Hamiltonian terms in the corresponding causal neighborhood of the parameterized gate can contribute to the gradient. For fixed $D$, this region has bounded size independent of $N$. Thus local-cost gradients in fixed-depth local circuits are governed by finite causal regions rather than by the full many-body Hilbert space. This is the locality distinction used in Sec.~VI, and it is different from the concentration-of-measure mechanism that can lead to barren plateaus in sufficiently random or sufficiently expressive circuits~\cite{mcclean2018barren,cerezo2021cost,holmes2022connecting}.

We next state the related restriction on the operator directions available at fixed depth. Let
$\mathcal{G}_0=\{G_1,G_2,\ldots\}$
be the set of local generators used by the circuit. If arbitrary depth is allowed, these generators may generate a larger Lie algebra through linear combinations and nested commutators:
\begin{equation}
\begin{aligned}
\mathfrak{g}
&=
\operatorname{Lie}(\mathcal{G}_0) \\
&=
\operatorname{span}\Big\{
G_j,\,
i[G_j,G_k],\,
i^2[G_j,[G_k,G_l]], \\
&\hspace{1.2cm}
i^3[G_j,[G_k,[G_l,G_m]]],
\ldots
\Big\}.
\end{aligned}
\end{equation}
The structure of this algebra is relevant to Lie-algebraic analyses of loss concentration and barren plateaus~\cite{ragone2024lie}. At fixed local depth, however, the circuit accesses only those directions that can be generated inside finite causal regions. We denote this fixed-depth accessible span by $\mathcal{A}_D$.

\noindent\textit{Proposition 2. Restricted finite-depth accessible span.}
For fixed local depth $D=O(1)$, bounded gate range, and bounded lattice coordination number,
\begin{equation}
\dim \mathcal{A}_D
\leq
c(D)N,
\end{equation}
where $c(D)$ depends on the depth, geometry, and local gate structure, but not on the total system size $N$.

\noindent\textit{Proof.}
At fixed depth, a local operator direction can spread only through overlapping local gates. If two operators have disjoint support, then
\begin{equation}
{\rm supp}(A)\cap{\rm supp}(B)=\emptyset
\quad \Rightarrow \quad
[A,B]=0 .
\end{equation}
Thus nested commutators can enlarge support only along paths contained in a finite causal region.

Let $q_D$ be the maximum number of qubits in such a region. For fixed $D$, bounded gate range, and bounded coordination number, $q_D$ is independent of $N$. The operator space supported on one region has dimension at most $4^{q_D}-1$, and the number of possible causal regions grows at most linearly with the number of lattice sites. Hence
\begin{equation}
\dim \mathcal{A}_D
\leq
N(4^{q_D}-1)
\equiv
c(D)N,
\end{equation}
with $c(D)$ independent of $N$ at fixed depth.

This statement applies only to the fixed-local-depth setting described above. It does not describe the full algebra generated by the same gates at arbitrary depth. By contrast, the full traceless operator space on $N$ qubits has dimension $4^N-1$. Thus, the fixed-depth restriction does not concern the generators themselves, but the finite causal regions through which they can act in a local circuit.

\section{Reference values of the inverse participation ratio} \label{app:ipr}

For a pure state expanded in the computational basis,
\begin{equation}
\ket{\psi}=\sum_{z=1}^{D} c_z \ket{z},
\qquad
\sum_{z=1}^{D}|c_z|^2=1,
\end{equation}
we define the inverse participation ratio (IPR) as
\begin{equation}
\mathrm{IPR}=D\sum_{z=1}^{D}|c_z|^4,
\qquad D=2^N.
\end{equation}
This normalization is convenient because it assigns distinct reference values to several limiting cases.

For a computational-basis product state, all weight is concentrated on a single basis vector, so that
\begin{equation}
\mathrm{IPR}=D.
\end{equation}
Thus, large IPR indicates strong localization in the computational basis.

For a perfectly uniform equal-amplitude state, with $|c_z|^2=1/D$ for all $z$, one instead finds
\begin{equation}
\mathrm{IPR}=D \sum_{z=1}^{D}\frac{1}{D^2}=1.
\end{equation}
This is the minimum possible value under the present normalization.

A different benchmark is provided by Haar-random pure states. In that case the computational-basis probabilities $p_z=|c_z|^2$ follow the Porter-Thomas form in the large-$D$ limit~\cite{arute2019quantum,zlokapa2023boundaries}, and their second moment is
\begin{equation}
\mathbb{E}[p_z^2]=\frac{2}{D(D+1)}.
\end{equation}
Therefore,
\begin{equation}
\mathbb{E}[\mathrm{IPR}]
=
D\sum_{z=1}^{D}\mathbb{E}[p_z^2]
=
\frac{2D}{D+1},
\end{equation}
which approaches
\begin{equation}
\mathbb{E}[\mathrm{IPR}] \to 2
\qquad
(D\to\infty).
\end{equation}

Hence, within this normalization, the IPR distinguishes three regimes:
strong localization ($\mathrm{IPR}\gg 1$), perfectly uniform delocalization ($\mathrm{IPR}=1$), and random-state-like delocalization ($\mathrm{IPR}\approx 2$). We emphasize that $\mathrm{IPR}\approx 2$ reflects only the statistics of computational-basis amplitudes and does not by itself imply Haar randomness or the formation of a unitary $2$-design.

\FloatBarrier
\section{Additional Cluster-Ising VQE Results} \label{app:cluster-vqe-additional}

\begin{figure}[H]
    \centering
    \includegraphics[width=1.0\columnwidth]{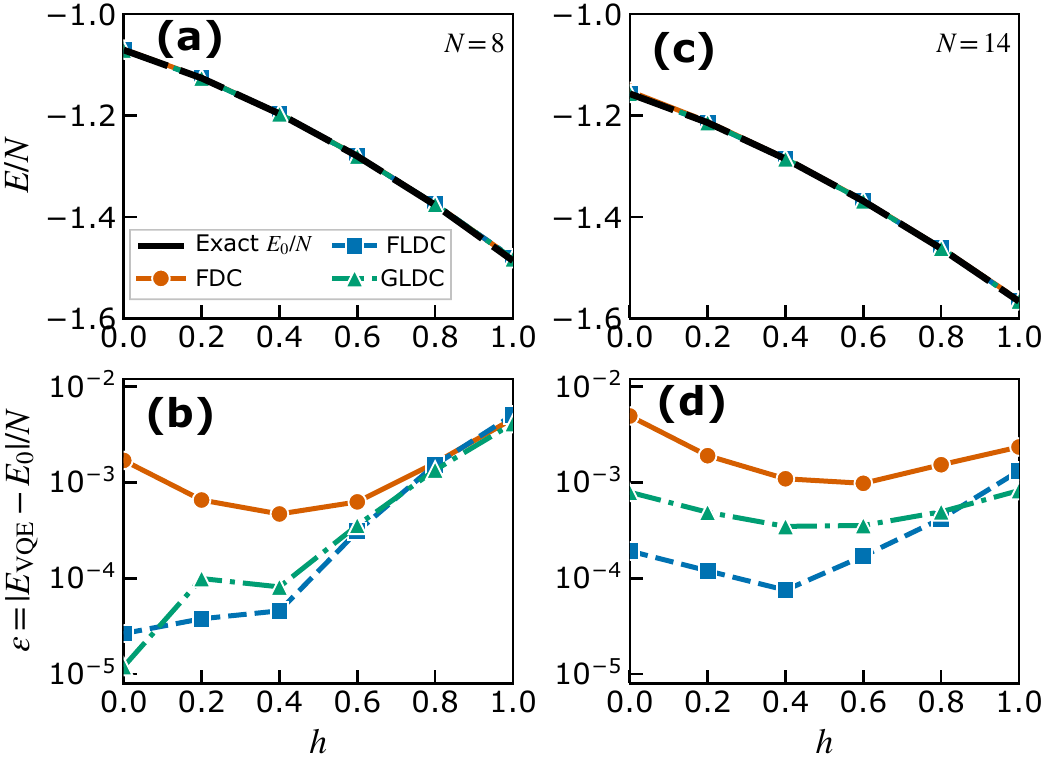}
    \caption{
    Additional Cluster-Ising VQE results for $N=8$ [panels (a) and (b)] and $N=14$ [panels (c) and (d)]. The upper panels show the optimized energy per qubit, while the lower panels show
    $\epsilon=|E_{\rm VQE}-E_0|/N$. The VQE protocol is the same as in Fig.~\ref{fig:CI_VQE}.}
    \label{fig:CI_VQE_appendix}
\end{figure}

Figure~\ref{fig:CI_VQE_appendix} collects supplementary Cluster-Ising VQE results for $N=8$ and $N=14$. These additional system sizes are consistent with the main-text observations for the representative sizes $N=12$ and $N=16$: the energy-per-qubit curves remain close to the exact benchmark, while the normalized error provides a more sensitive view of the ansatz-dependent differences.

\section{Full Cluster-Ising Statistical Diagnostics} \label{app:cluster-stats-full}

\begin{figure*}[!t]
    \centering
    \includegraphics[width=0.9\textwidth]{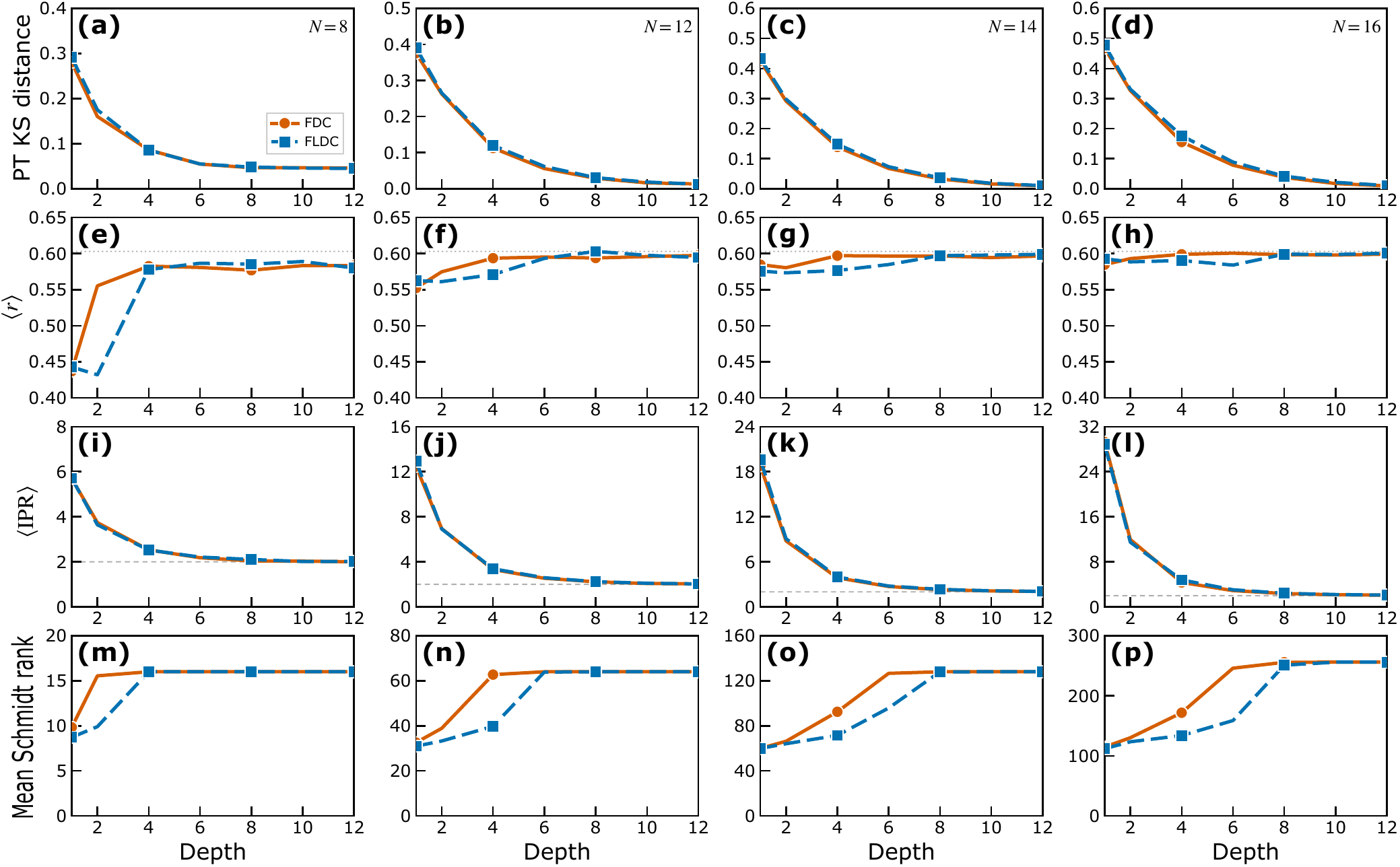}
    \caption{
    Full cluster-Ising statistics-only diagnostics for $N=8,12,14$, and $16$. Rows show the Porter-Thomas KS distance, the entanglement-spectrum adjacent-gap ratio $\langle r\rangle$, the normalized inverse participation ratio $\langle{\rm IPR}\rangle$, and the mean Schmidt rank versus circuit depth. In the $\langle r\rangle$ row, the horizontal line marks the GUE reference value. In the IPR row, the horizontal reference lines mark the random-state-like delocalization value $\langle{\rm IPR}\rangle\simeq 2$ for the normalization used here. Each point is averaged over 100 random circuit samples.
}
    \label{fig:CI_stats_appendix}
\end{figure*}

Figure~\ref{fig:CI_stats_appendix} shows the full cluster-Ising statistical diagnostics across $N=8$, $12$, $14$, and $16$. In addition to the $r$ ratio, Porter-Thomas distance, and inverse participation ratio shown in the main text, Fig.~\ref{fig:CI_stats_appendix} also includes the mean Schmidt rank. Here the Schmidt rank refers to the rank of the reduced density matrix $(\rho_A=\mathrm{Tr}_B(|\psi\rangle\langle\psi|))$ for the same bipartition used in the entanglement-spectrum calculation, equivalently the number of nonzero eigenvalues of $(\rho_A)$. The same qualitative depth dependence is visible across the full set of sizes studied.

\section{Additional toric-code data for the U-shaped ordering}
\label{app:toric-u-results}

We also repeated the generalized toric-code calculations using the U-shaped within-plaquette ordering motivated by Ref.~\cite{zhang2024absence}. The Hamiltonian, ansatz definitions, optimization procedure, and statistical
diagnostics are the same as in the main text.

\begin{figure*}[ht]
    \centering
    \includegraphics[width=0.85\columnwidth]{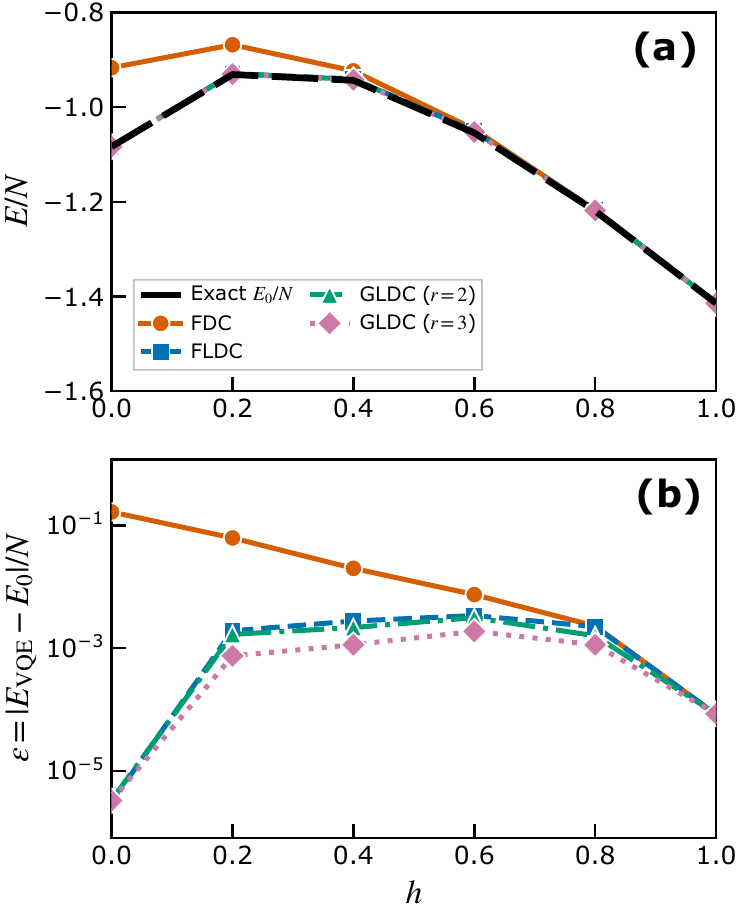}
    \caption{
    Generalized toric-code VQE results for the $N=12$ edge-qubit geometry with U-shaped within-plaquette ordering.
    Panel (a) shows the optimized energy per qubit, and panel (b) shows the normalized error
    $\epsilon=|E_{\rm VQE}-E_0|/N$.
    The VQE protocol is the same as in Fig.~\ref{fig:TC_VQE}.
}
    \label{fig:app_toricU_vqe}
\end{figure*}

Figure~\ref{fig:app_toricU_vqe} shows the VQE results for this ordering. The optimized energies track the exact result over the field range, and the normalized error shows the same overall pattern seen in the main toric-code
data: the deeper grouped circuits and FLDC improve the low-field accuracy relative to the FDC circuit.

\begin{figure*}[ht]
    \centering
    \includegraphics[width=0.90\textwidth]{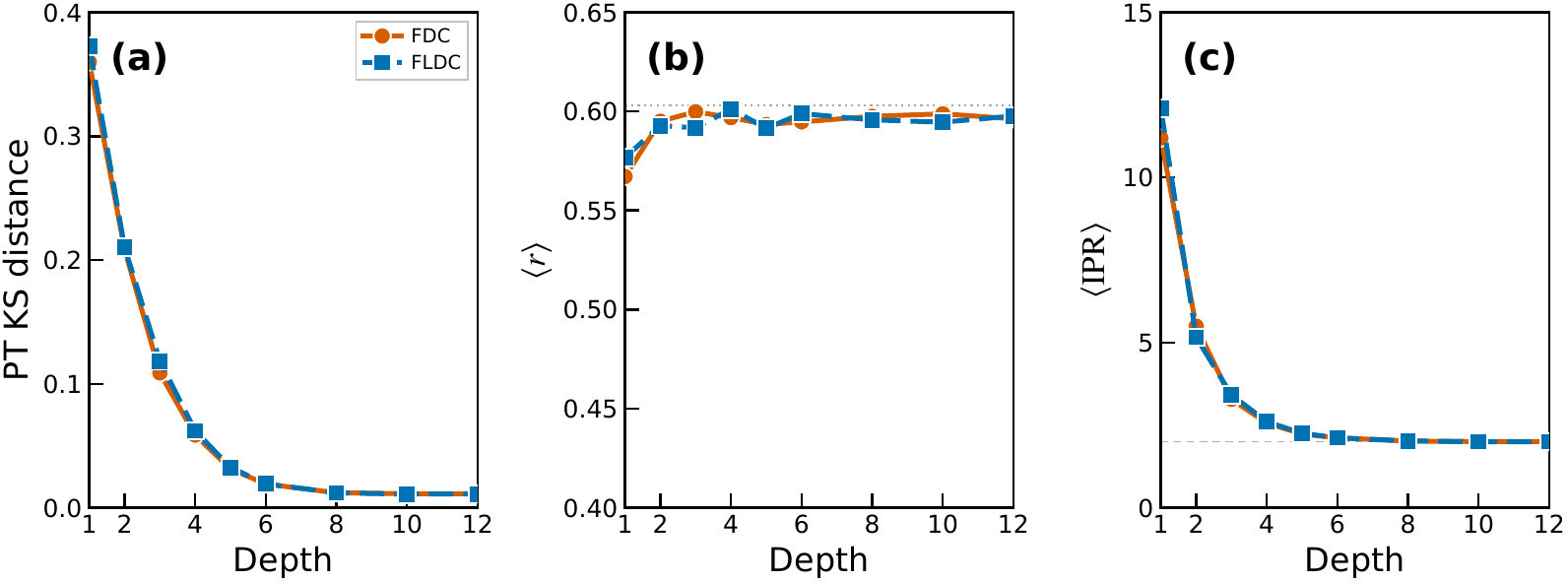}
    \caption{
    Statistics-only diagnostics for the generalized toric-code ansatz with U-shaped within-plaquette ordering. The panels show (a) the Porter-Thomas KS distance, (b) the entanglement-spectrum adjacent-gap ratio $\langle r\rangle$, and (c) the normalized inverse participation ratio $\langle{\rm IPR}\rangle$. In panel (b), the dotted horizontal line marks the GUE reference value. In panel (c), the horizontal reference line marks the random-state-like delocalization value $\langle{\rm IPR}\rangle\simeq 2$ for the normalization used here. Each point is averaged over 200 random circuit samples.}
    \label{fig:app_toricU_stats}
\end{figure*}

The corresponding statistics-only results are shown in Fig.~\ref{fig:app_toricU_stats}. The trends are very similar to the claw-ordering case discussed in the main text. The value of $\langle r\rangle$ moves toward the GUE reference, the Porter-Thomas KS distance decreases with depth, and the normalized IPR moves toward the random-state-like value discussed in Appendix~\ref{app:ipr}. Therefore, the U-shaped ordering gives the same qualitative picture and does not change the main conclusion from the toric-code data.

\section{Additional Gradient Diagnostics}
\label{app:additional_gradient_diagnostics}

This appendix gives an additional view of the gradient data used to support the trainability discussion in Sec.~VI. The mean per-parameter gradient variance is shown in the main text in Fig.~\ref{fig:grad_var_vs_N}, where it is organized as a function of system size for several fixed circuit depths. Here we report the averaged squared gradient norm. This quantity is a useful finite-size diagnostic, but it should not be interpreted as an asymptotic scaling proof.

\begin{figure*}[!t]
\centering
\includegraphics[width=0.95\textwidth]{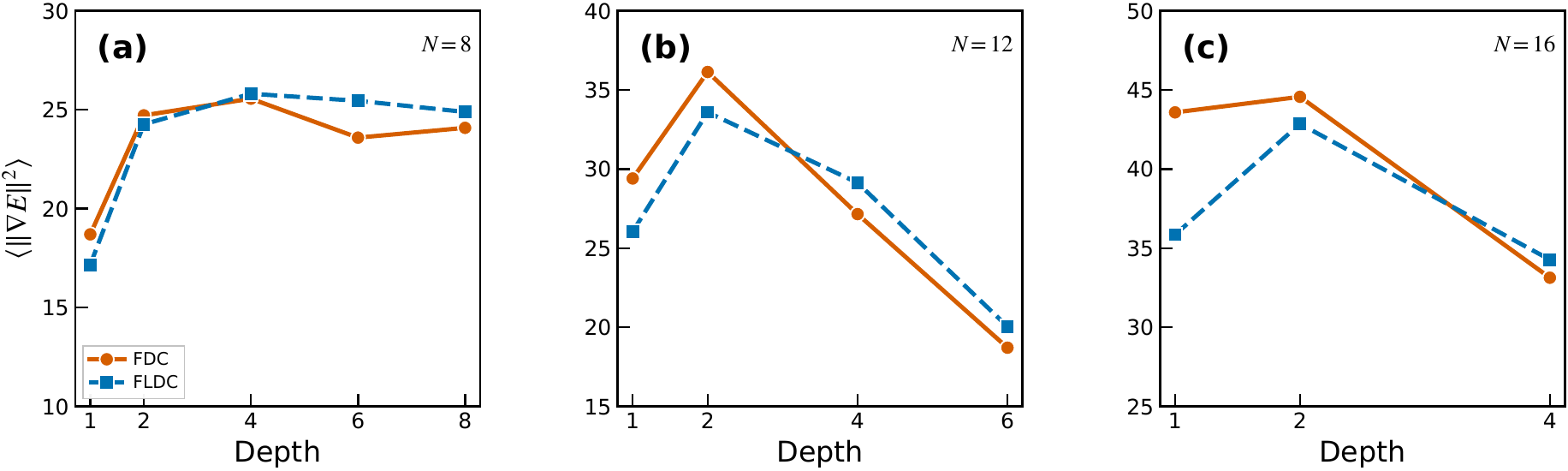}
\caption{
Averaged squared gradient norm for the cluster-Ising model at field strength $h=0.6$, with $J=J'=1$, shown as a function of circuit depth. Panels (a)--(c) correspond to $N=8$, $N=12$, and $N=16$, respectively, with FDC and FLDC shown in each panel. Each point is estimated from 100 random parameter samples.}
\label{fig:gradient_norm_sq_depth_appendix}
\end{figure*}

Figure~\ref{fig:gradient_norm_sq_depth_appendix} shows the averaged squared gradient norm,
$(\langle |\nabla E|^2\rangle)$, as a function of depth. This quantity combines the contributions from all trainable parameters and therefore gives a complementary measure of the total gradient signal. Unlike the mean per-parameter variance, however, it also depends on the number of parameters in the circuit, so we do not use it by itself as a barren-plateau diagnostic. Instead, it provides an additional check that the total gradient signal does not collapse over the accessible depths. This is consistent with the variance data in Fig.~\ref{fig:grad_var_vs_N} and with the broader conclusion that the statistical signatures of complexity observed here do not, by themselves, imply a barren plateau in the studied finite-size regime.

\vspace*{14.0em}
\vspace{30.5em}


\bibliographystyle{apsrev4-2}
\hypersetup{allcolors=black}
\bibliography{reference}

\end{document}